\documentclass{article}
\usepackage{spconf,amsmath,graphicx,wrapfig,bm}
\usepackage{amsfonts,amssymb}

\title{Exploring the effects of channel sparsity on neural network pruning for acoustic scene classification}
\name{Yiqiang Cai, Shengchen Li}
\address{School of Advanced Technology\\
Xi'an Jiaotong-Liverpool University\\
111 Ren'ai Road, Suzhou, China }
\begin{document}
%
\maketitle
\begin{abstract}
Acoustic Scene Classification (ASC) algorithms are usually expected to be deployed in resource-constrained systems. Existing works reduce the complexity of ASC algorithms by pruning some components, e.g. pruning channels in neural network. In practice, neural networks are often trained with sparsification such that unimportant channels can be found and further pruned. However, little efforts have been made to explore the the impact of channel sparsity on neural network pruning. To fully utilize the benefits of pruning for ASC, and to make sure the model performs consistently, we need a more profound comprehension of channel sparsification and its effects. This paper examines the internal weights acquired by convolutional neural networks that will undergone pruning. The study discusses how these weights can be utilized to create a novel metric, \textit{Weight Skewness (WS)}, for quantifying the sparsity of channels. We also provide a new approach to compare the performance of different pruning methods, which balances the trade-off between accuracy and complexity. The experiment results demonstrate that 1) applying higher channel sparsity to models can achieve greater compression rates while maintaining acceptable levels of accuracy; 2) the selection of pruning method has little influence on result 1); 3) MobileNets exhibit more significant benefits from channel sparsification than VGGNets and ResNets.
\end{abstract}
\begin{keywords}
Channel sparsity, neural network pruning, acoustic scene classification
\end{keywords}

\begin{figure*}[ht]
\centering
\includegraphics[width=1\textwidth]{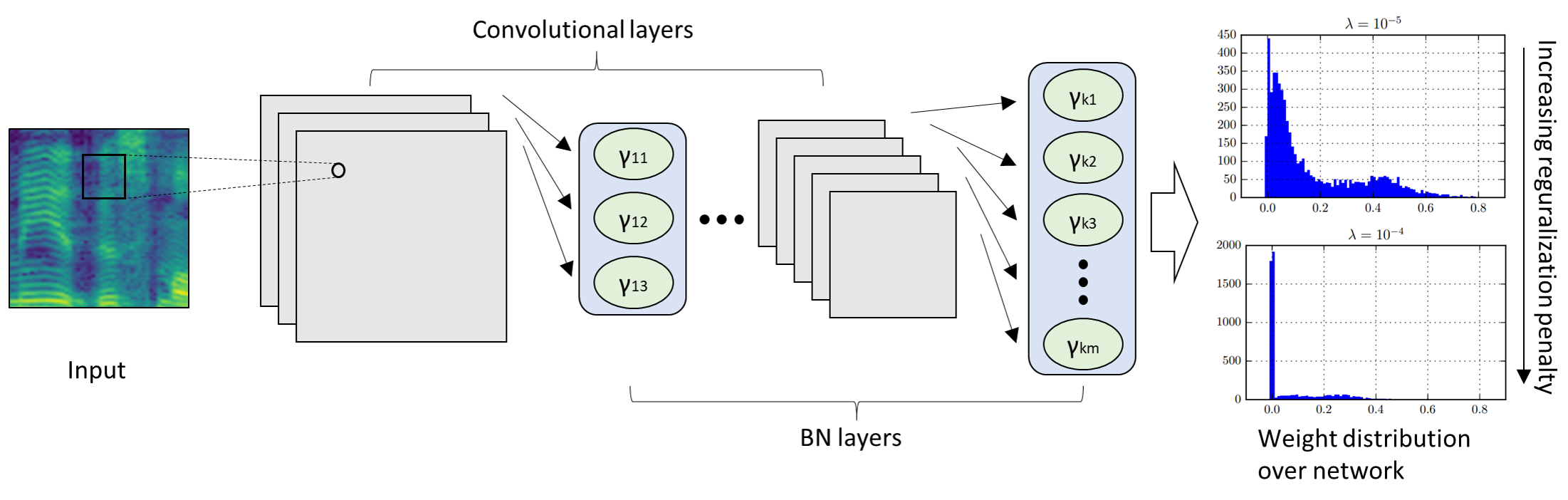}
\caption{\label{fig:sparsification}Sparsifying a CNN-based ASC system for channel pruning. The bar charts indicate the weight distribution of scaling factors $\gamma$ in BN layers over network.}
\end{figure*}

\section{Introduction}
\label{sec:intro}

Acoustic scene classification (ASC) \cite{barchiesi2015acoustic} is the task of categorizing an audio signal into one of several predefined classes based on the environmental sounds present in the recording. It has become an important research topic due to its wide range of applications. However, the conventional ASC methods typically require significant computational resources, which limits the practical application in resource-constrained scenarios, such as in mobile devices, internet of things (IoT) devices, and embedded systems. There has been a growing development for efficient and accurate ASC algorithms that can be deployed in resource-constrained platforms, like microphone \cite{7846603}, headphone \cite{9673469} and audiphone \cite{9182160}.

Several approaches have been introduced to address the issue of low-complexity ASC, including low rank approximation \cite{jaderberg2014speeding}, weight quantization \cite{chen2015compressing}, neural network pruning \cite{lecun1989optimal} and knowledge distillation \cite{hinton2015distilling}. These methods aim to reduce the computational cost, limit the number of parameters in the model, or both, while still maintaining a high classification accuracy. Among methods mentioned, neural network pruning is widely used, which prunes the components of a neural network with little contribution to the final performance. Pruning methods can be broadly divided into two subcategories based on the type of components being pruned, unstructured pruning and structured pruning. Unstructured pruning \cite{han2015learning}, so called weight pruning at first, prunes the sole weight in the scope of whole network, but usually requires dedicated hardware. By contrast, structured pruning \cite{wen2016learning} avoids the problem by pruning the structured components of a neural network. Recently, channel pruning, which involves removing entire channels (i.e. feature maps) from the network, has been demonstrated as an effective approach for achieving model compression by many researchers \cite{he2017channel} \cite{guo2020channel} \cite{yang2022channel}. Specially, Liu et al. \cite{liu2017learning} and Ye et al. \cite{ye2018rethinking} induce sparsity on the channels of neural network before pruning and achieve large compression rate without sacrificing accuracy.

However, the impact of channel sparsity on neural network pruning has not been investigated. While previous studies \cite{singh2019deep} \cite{wang2019svd} have demonstrated the effectiveness of channel pruning for ASC, it still remains unclear how the degree of channel sparsity affects the performance of neural network pruning.

In this paper, our goal is to offer a new perspective on channel pruning to researchers and practitioners who are interested in optimizing ASC systems for resource-constrained environments. Firstly, we propose a novel metric, \textit{Weight Skewness (WS)}, to measure the channel sparsity of neural networks. Additionally, a new approach, \textit{Pruning Knee (PK)}, is also introduced for evaluating the performance of pruning techniques that considers the balance between accuracy and complexity. In the experiment, we implement 3 CNN architectures, and train networks with different degree of regularization so as to get various \textit{WS}. Next, channel pruning is performed to discover the \textit{PK}. Furthermore, linear regression and Pearson correlation is used to examine the relationship between \textit{WS} and \textit{PK}. In addition, the effects of pruning methods and hyper-parameters of a single network architecture on this relationship are explored. Consequently, we draw a conclusion based on the experimental results. The main contributions can be concluded as follows:

1) We propose a novel metric to measure the sparsity of channels in neural networks for ASC, namely \textit{Weight Skewness (WS)}. Different from previous works related to channel pruning, we are interested in studying the properties of channel pruning rather than putting forward another channel pruning method.

2) We suggest a new approach to examine the performance of neural network pruning, which strikes a balance in the trade-off between model complexity and accuracy.

3) We analyze the impact of channel sparsity on neural network pruning by exploring the relationship between \textit{WS} and \textit{PK}, and demonstrate a strong positive linear correlation between channel sparsity and pruning performance.

The paper is organinzed as below. The Section \ref{sec:method} introduces the methods we use to obtain and analyze \textit{WS} and \textit{PK}. Technical details of experiments are elaborated in Section \ref{sec:exp}. Section \ref{sec:result} sets forth the experimental results of the study. Section \ref{sec:discussion} interprets and analyzes the results. Conclusion is drawn in Section \ref{sec:conclusion}.

\section{Methods}
\label{sec:method}

\subsection{Channel Sparsification}
\label{ssec:sparsification}
Channel sparsification aims to identify the redundant or unimportant channels (i.e., feature maps) in a convolutional neural network (CNN). The main challenge is how to distinguish the unimportant channels from useful channels.

In recent years, several channel pruning methods \cite{liu2017learning} \cite{ye2018rethinking} have been proposed that leverage the scaling factors in batch normalization (BN) \cite{ioffe2015batch} layers to guide the channel sparsification process, which achieved significant compression rates while preserving or even improving the performance of CNNs. BN is a popular technique to normalize the activations of CNNs, and it introduces learnable scaling factors to restore the representation power of the network.

The schematic overview of scaling factor-based channel sparsification is presented in Fig. \ref{fig:sparsification}. Suppose a CNN with $k$ convolutional layers $\{l_1,...,l_k\}$ and $m$ filters for each layer $\{f_1,...,f_m\}$, and each convolutional layer is followed by a BN layer. The transformation of $i_{th}$ BN layer can be denoted by Eq. (\ref{eq:bn1}) and (\ref{eq:bn2}),

    \begin{equation}
    \label{eq:bn1}
        \hat{x_i} = \frac{x_{i}-\mu_{c}^{(B)}}{\sqrt{\sigma_{c}^{2(B)}+\epsilon}}
    \end{equation}
    \begin{equation}
    \label{eq:bn2}
        x_{i+1} = \gamma \cdot \hat{x_i} + \beta
    \end{equation}

\noindent
where $x_{i}$ and $x_{i+1}$ respectively represent the input and output of the BN layer, $\mu_{c}^{(B)}$ and $\sigma_{c}^{(B)}$ separately indicate the mean and standard deviation of input on the scale of a mini-batch B, $\epsilon$ is an extremely small constant for numerical stability, $\gamma$ and $\beta$ are trainable parameters for scaling and shifting. By exploiting the magnitude of these scaling factors $\gamma$, channel pruning methods can identify and remove the channels that contribute less to the final output of the network. The bar chart of Fig. \ref{fig:sparsification} presents the weight distribution of scaling factors $\gamma$ over network. Channels with smaller scaling factors contribute less to the output of the BN layer, and thus can be pruned without significantly affecting the performance of the network. In other words, the scaling factors $\gamma$ can be directly used to measure the significance of channels.

During training, adding regularization to the scaling factor $\gamma$ leads to channel sparsification as shown in Eq. (\ref{eq:loss}),

    \begin{equation}
    \label{eq:loss}
        \min_{\theta} \sum_{i=1}^n L(f(\theta;x_i),y_i) + \lambda R(\gamma)
    \end{equation}

\noindent
where $L(\cdot)$ is the loss function of neural network $f(\cdot)$ , $x_i$ and $y_i$ respectively denote the input feature and label of the $i_{th}$ sample of the training set ${\{(x_i,y_i)\}}^n_{i=1}$, $\theta$ represent the parameters of network, $\lambda$ is a hyper-parameter for adjusting the degree of regularization. The penalty term $\lambda R(\gamma)$ forces the value of $\gamma$ to near zero during training as shown in Fig. \ref{fig:sparsification}. Meanwhile, a network may achieve different sparsities by controlling $\lambda$ during training.

\subsection{Measuring Channel Sparsity}
\label{ssec:sc}
During the process of channel sparsification, the more number of scaling factors $\gamma$ closing to zero indicates a sparser network. As shown in the bar chart of Fig. \ref{fig:sparsification}, the weight distribution of scaling factors $\gamma$ over network is excessively shifted to the left side. Due to the skewed shape of weight distribution, we adopt \textit{skewness} to propose a statistical metric, \textit{Weight Skewness (WS)}, for measuring the channel sparsity in CNN. Consider the scaling factors $\gamma$ over network as $\{\gamma_{11},...,\gamma_{ij},...,\gamma_{km}\}$, we get:

\begin{equation}
\label{eq:ws1}
    WS= \frac{\sum^k_{i=1}\sum^m_{j=1}(\gamma_{ij}-\overline{\gamma})^3}{(k\cdot m-1)\cdot\sigma^3}.
\end{equation}
\begin{equation}
\label{eq:ws2}
    \overline{\gamma}=\frac{1}{k\cdot m}\sum^k_{i=1}\sum^m_{j=1}\gamma_{ij}
\end{equation}
\begin{equation}
\label{eq:ws3}
    \sigma=\sqrt{\frac{\sum^k_{i=1}\sum^m_{j=1}(\gamma_{ij}-\overline{\gamma})^2}{k\cdot m}}
\end{equation}

As the tail of the weight distribution of scaling factors $\gamma$ over network extends to the right side, \textit{WS} is generally greater than 0, and the value of \textit{WS} is expected to increase as the channels getting sparser.

\subsection{Prune Knee}
\label{ssec:knee}
\begin{figure}
\centering
\includegraphics[width=0.4\textwidth]{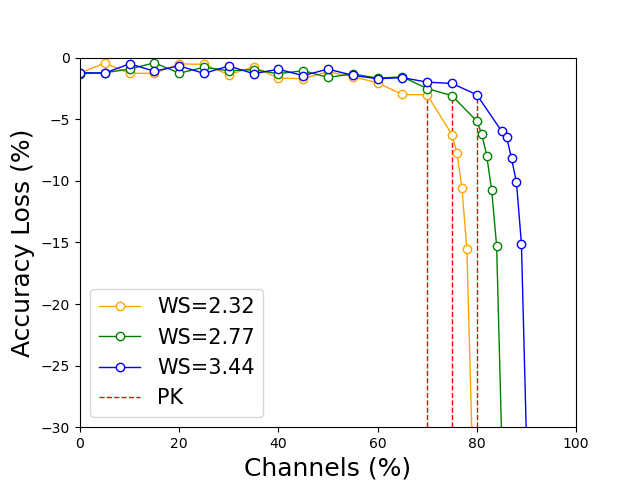}
\caption{\label{fig:knee}Detecting the \textit{Prune Knee (PK)} by using \textit{Kneedle}. Examples are selected from experiment data of VGGNet(depth=16).}
\end{figure}

Although pruning helps to improve the efficiency of neural networks, the trade-off between performance and complexity in neural network pruning is not negligible \cite{han2015learning} \cite{li2016pruning}. For example, if a large number of network parameters are pruned, the remaining parameters may not be able to capture the full complexity of the input data, resulting in reduced performance as shown in Fig. \ref{fig:knee}. However, the effectiveness of a neural network pruning method is often evaluated based on accuracy and complexity separately, which cannot capture the trade-off between them.

Therefore, we propose a new approach, \textit{Prune Knee (PK)}, to evaluate the performance of neural network pruning in a more comprehensive way. \textit{PK} is defined as the balance point between accuracy loss and pruning rate of parameters. A larger \textit{PK} indicates the model has more parameters can be pruned without much harm to accuracy. Since the knee of a curve is commonly the point of maximum curvature, a knee detection technique, \textit{Kneedle} \cite{5961514}, is introduced in this paper to help determine the knee of pruning curves as show in Fig. \ref{fig:knee}. It works as follows, for a discrete data set $D=\{(x_i,y_i)\in\mathbb{R}|x_i,y_i\geq0\}$, the points $(x_i,y_i)$ are firstly min-max normalized. Secondly, the data set of difference $D_d=\{{(x_{d_i},y_{d_i})|x_{d_i}=x_i,y_{d_i}=y_i-x_i}\}$ is computed. Thirdly, the local maxima of the difference curve $(x_{lmx_i},y_{lmx_i})$ are discovered by $x_{lmx_i}=x_{d_i},y_{lmx_i}=y_{d_i}|y_{d_{i-1}}<y_{d_i},y_{d_{i+1}}<y_{d_i}$. Finally, the knee will be detected as long as $(x_{d_i},y_{d_i})$ decreases blow the threshold $T_{lmx_i}$ before the next local maxima is reached. $T_{lmx_i}$ is calculated by Eq. (\ref{eq:knee}), where $\psi$ is a parameter to control the sensitivity.

\begin{equation}
\label{eq:knee}
    T_{lmx_i} = y_{lmx_i} - \psi\cdot\frac{\sum_{i=1}^{n-1}(x_{i+1}-x_n)}{n-1}
\end{equation}

\begin{table*}[t]
\centering
\begin{tabular}{c c|c c c|c c c}
\hline
\multicolumn{2}{c|}{unpruned}& \multicolumn{3}{c|}{pruned (Liu et al. \cite{liu2017learning})}& \multicolumn{3}{c}{pruned (Ye et al. \cite{ye2018rethinking})} \\
acc.& \textit{WS}& acc. loss& compr. rate& \textit{PK}& acc. loss& compr. rate& \textit{PK} \\
\hline \hline
$90.2$& $0.10$& $-2.2$& $67\%$& $61\%$& $-2.8$& $73\%$& $70\%$ \\
$89.4$& $0.26$& $-2.0$& $79\%$& $70\%$& $-2.0$& $79\%$& $75\%$ \\
$90.3$& $1.10$& $-3.3$& $81\%$& $76\%$& $-3.7$& $84\%$& $81\%$ \\
$91.3$& $1.79$& $-3.2$& $83\%$& $78\%$& $-3.1$& $86\%$& $84\%$ \\
$91.0$& $2.77$& $-3.4$& $86\%$& $81\%$& $-3.7$& $89\%$& $88\%$ \\
$93.5$& $3.44$& $-4.3$& $87\%$& $86\%$& $-4.5$& $89\%$& $89\%$ \\
\hline
\end{tabular}
\caption{\label{tab:dataset}Discrete data set of \textit{Weight Skewness (WS)} and \textit{Prune Knee (PK)}. Examples are selected from experiment data of VggNet (depth=16). Compression rate = the number of parameters of pruned model / the number of parameters of original model.}
\end{table*}

\section{Experiments}
\label{sec:exp}

The experiment is designed to examine the relationship between \textit{WS} and \textit{PK}. It has been divided into 3 steps. \textbf{1) Training sparse networks.} By controlling the hyper-parameter $\lambda$ in Eq. (\ref{eq:loss}) at the training process, we are able to obtain networks with different degree of channels sparsity and calculate \textit{WS} by Eq. (\ref{eq:ws1}). \textbf{2) Pruning.} Firstly, a trained model is iteratively pruned  with a small fixed pruning rate of channels for each step. Then, we get a discrete data set of pruning rate and accuracy loss after evaluations. Finally, the knee is detected from the data set by using \textit{Kneedle} \cite{5961514} and corresponding \textit{PK} is found. \textbf{3) Linear correlation analysis.} The linear regression and Pearson correlation are leveraged to reveal the relationship between \textit{WS} and \textit{PK}. 

\subsection{Dataset and Preprocessing} 
The experiment is designed for the task of Acoustic Scene Classification (ASC). The TAU Urban Acoustic Scenes 2020 Mobile development dataset \cite{Mesaros2018_DCASE} consists of audio recordings from 12 European cities in 10 acoustic scenes by using 4 devices. We collect the data captured by device A, which is a professional equipment for audio recording. The dataset condtains 14,400 segments, totally about 40 hours, and each segment has the same length of 10s and is produced in single-channel 44.1kHz 24-bit format. Moreover, we split the dataset into 2 subset, respectively 70$\%$ for training and 30$\%$ for validation. The audios are resampled by 44.1Hz and the features are extracted by using log mel-band energies with 40 bands, window length of 40ms and hop size of 62.5$\%$.

Data augmentation techniques play a crucial role in audio processing tasks, with the advantages of improving generalization abilities and reducing the effect of overfitting. We use Mixup \cite{zhang2017mixup} with $\alpha=0.4$ and Specaugmentation \cite{Park_2019} with a frequency mask of 4 and a time mask of 40 in all experiments.

\begin{table}
\centering
\begin{tabular}{c c|c c c c}
\hline
network& hyper-param& v1& v2& v3& v4 \\
\hline \hline
VGGNets& depth& 11& 13& 16& 19 \\
ResNets& depth& 11& 20& 29& 38 \\
MobileNets& width& 0.25& 0.50& 0.75& 1.00 \\
\hline
\end{tabular}
\caption{\label{tab:models}Summary of network architectures. }
\end{table}

\subsection{Network Architectures} 
As Tab \ref{tab:models} shows, three representative CNN architectures are selected in the experiment and v1-v4 indicate different variants for a single network. For VGGNet \cite{vgg}, we choose 4 variants with respectively 11, 13, l6 and 19 layers of depth. For ResNet \cite{he2016deep}, 4 variants with respectively 11, 20, 29 and 38 layers of depth are selected. For MobileNet \cite{howard2017mobilenets}, 4 variants with different widths are chosen, separately 0.25, 0.5, 0.75 and 1. The scaling factors $\gamma$ in BN layers are initialized by 0.5 to get a better performance according to \cite{he2015delving} and \cite{zagoruyko2016wide}. Dropout layers of all networks are removed as data augmentation methods already provide strong regularization.

All networks follow the same training setups, for 100 epochs with batch size 32, using Adam optimizer with learning rate to 0.1, momentum to 0.9, weight decay to 0.0001. We achieve sparsity at the level of channels by adding a L1-norm term, i.e. $R(\gamma)={\|\gamma\|}_1$, on the scaling factors $\gamma$ to the loss function in Eq. (\ref{eq:loss}). In addition, the value of $\lambda$ in Eq. (\ref{eq:loss}) is adjusted before training, in order to get networks with various sparsities. It is worth mentioning if $\lambda$ is assigned a large value, the model will suffer a dramatic degeneration of performance, so sparse networks are trained with $\lambda$ from a tiny value to an upper limit (varies for different networks) for less than 3$\%$ accuracy loss comparing with the top performance. \textit{WS} is calculated once after each sparse training as shown in the left of Tab \ref{tab:dataset}, following Eq. (\ref{eq:ws1}). 

\begin{figure*}[t]
\centering
\includegraphics[width=0.8\textwidth]{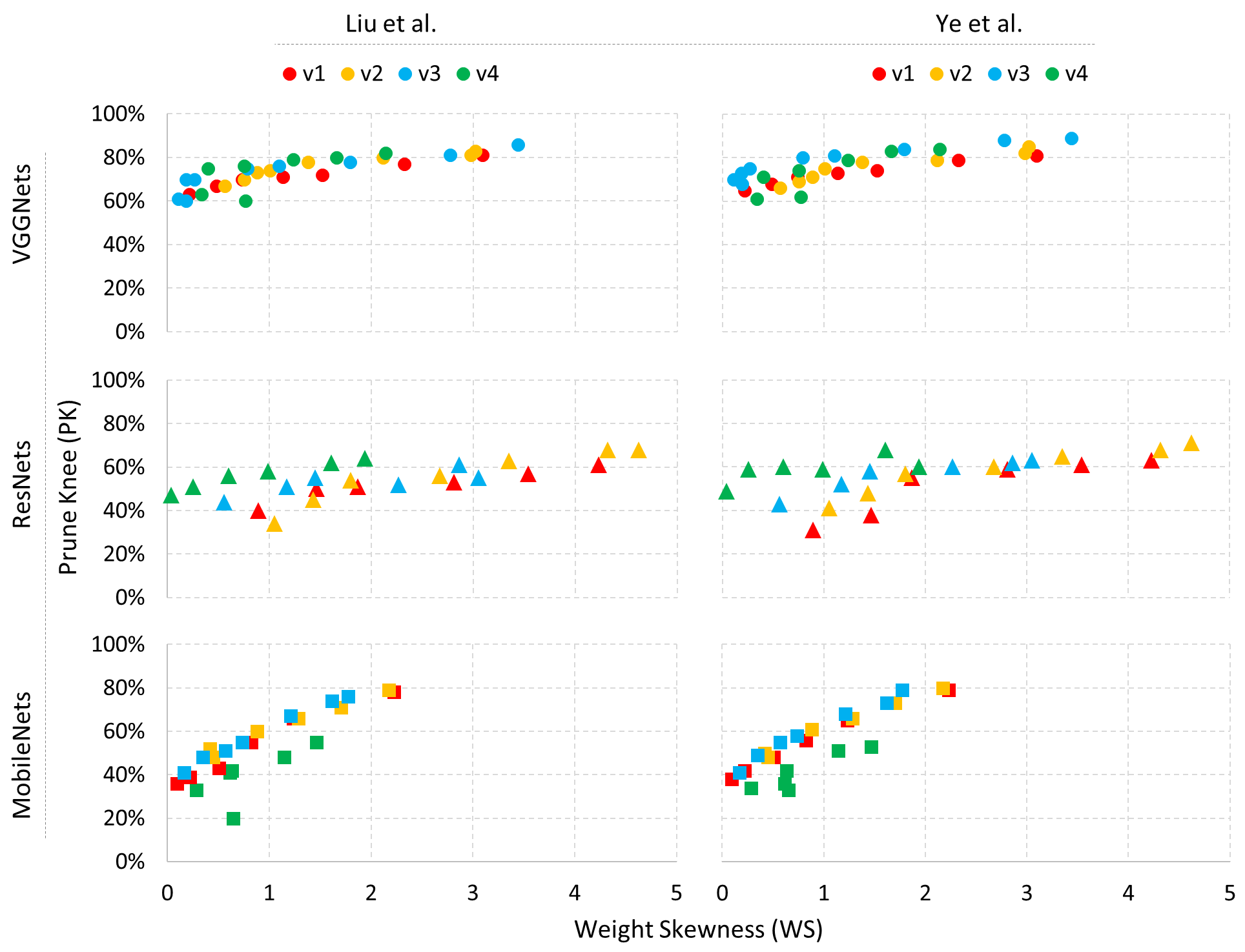}
\caption{\label{fig:result}\textbf{Scatter Plot of \textit{Weight Skewness (WS)} and \textit{Prune Knee (PK)}.}}
\end{figure*}

\subsection{Pruning}
Two scaling factor-based channel pruning methods are selected for comparison in the experiment. Liu et al. \cite{liu2017learning} trained sparse network by introducing Lasso regularization on scaling factors and removed the redundant channels on the global scale. Ye et al. \cite{ye2018rethinking} designed an iterative shrinking algorithm to update the scaling factors and trim unimportant channels accordingly. These two pruning methods are performed separately after the training process. Besides, we also fine-tune the model for 5 epochs \cite{han2015learning} at training set after pruning with the same optimizer as in training process, to recover some performance loss.

To be specific, the channels of trained models are iteratively pruned for a 5$\%$ pruning rate, starting from the channels with lowest values of scaling factors $\gamma$. Meanwhile, the performance of the pruned model is evaluated after each pruning iteration, so as to obtain the discrete data set of pruning rate and accuracy loss. Then, the knee of pruning curve is detected by using Kneedle \cite{5961514} and corresponding \textit{PK} is revealed as shown in Tab \ref{tab:dataset}. It is worth mentioning that we repeat this approach for each CNN and observe the performance decrease of all CNNs at \textit{PK} is approximately between 2$\%$ and 5$\%$.

\subsection{Evaluation}
The impact of channel sparsity on neural network pruning is evaluated by comparing the relationship between \textit{WS} and \textit{PK} over different CNN architectures. The analysis tools include linear regression and Pearson correlation. 

After training and pruning in the experiment, we get two discrete data set, $\{\textit{WS}_i\}^n_{i=1}$ and $\{\textit{PK}_i\}^n_{i=1}$. The linear regression analysis between \textit{WS} and \textit{PK} can be denoted by Eq. (\ref{eq:m}) and (\ref{eq:b}), and Pearson correlation coefficients is calulated by Eq. (\ref{eq:pearson}),
    \begin{equation}
    \label{eq:m}
        m=\frac{\sum^n_{i=1}({\textit{WS}}_i-\overline{\textit{WS}})({\textit{PK}}_i-\overline{\textit{PK}})}{\sum^n_{i=1}({\textit{WS}}_i-\overline{{\textit{WS}}_i})^2}
    \end{equation}
    
    \begin{equation}
    \label{eq:b}
        b=\overline{\textit{PK}}-m\cdot\overline{\textit{WS}}
    \end{equation}
    
    \begin{equation}
    \label{eq:pearson}
        r_{\textit{WS,PK}}=\frac{\sum_{i=1}^{n}(\textit{WS}_i-\overline{\textit{WS}})(\textit{PK}_i-\overline{\textit{PK}})}{\sqrt{\sum_{i=1}^{n}(\textit{WS}_i-\overline{\textit{WS}})^2}\sqrt{\sum_{i=1}^{n}(\textit{PK}_i-\overline{\textit{PK}})^2}}
    \end{equation}

\noindent
where $m$ and $b$ are slope and bias of the linear equation $\textit{PK}=m\cdot \textit{WS}+b$. $\overline{\textit{WS}}$ and $\overline{\textit{PK}}$ represent the sample means of \textit{WS} and \textit{PK} respectively. $r_{\textit{WS,PK}}$ represents the Pearson correlation coefficient between \textit{WS} and \textit{PK}. As show in Eq. (\ref{eq:pearson}), the numerator represents the covariance between \textit{WS} and \textit{PK}, while the denominator represents the product of their standard deviations. 

\section{Results}
\label{sec:result}
The experiments are conducted to explore the influence of channel sparsity on neural network pruning by investigating the relationship between \textit{WS} and \textit{PK}. The data set are collected by using 2 channel pruning methods from 3 CNN architectures, 4 variants for each CNN. We then performed linear regression analysis and computed Pearson correlation coefficients to examine the relationship between \textit{WS} and \textit{PK}.

Fig. \ref{fig:result} shows the comparative plots between \textit{WS} and \textit{PK}. The shapes of plots represent different CNN architectures, circle for VGGNets, triangle for ResNets and square for MobileNets. In addition, the colors of plots indicate different variants of a single CNN architecture, increasing depth or width from red to green. It can be easily seen that \textit{PK} grows with the increase of \textit{WS} in most cases and the scattered points are relatively concentrated on a line.

    \begin{table*}[t]
    \centering
    \begin{tabular}{c|c c c c|c c c c}
    \hline
    & \multicolumn{4}{|c}{Liu et al. \cite{liu2017learning}}& \multicolumn{4}{|c}{Ye et al. \cite{ye2018rethinking}} \\
    network& v1& v2& v3& v4& v1& v2& v3& v4 \\
    \hline \hline
    VGGNets& \textbf{0.979}& 0.922& 0.885& 0.687& 0.976& 0.931& 0.928& 0.847 \\
    ResNets& 0.939& 0.938& 0.799& \textbf{0.973}& 0.896& 0.949& 0.922& 0.690 \\
    MobileNets& 0.723& 0.980& 0.988& 0.989& 0.912& 0.991& 0.990& \textbf{0.996} \\
    \hline
    \end{tabular}
    \caption{\label{tab:pearson}Pearson correlation coefficients of \textit{Weight Skewness (WS)} and \textit{Prune Knee (PK)}. $r_{\textit{WS,PK}}$ can be solved by Equation (\ref{eq:pearson}). The data in bold highlights the highest value for each network architecture.}
    \end{table*}

    \begin{table*}[t]
    \centering
    \begin{tabular}{c c|c c c c|c c c c}
    \hline
    & & \multicolumn{4}{|c}{Liu et al. \cite{liu2017learning}}& \multicolumn{4}{|c}{Ye et al. \cite{ye2018rethinking}} \\
    network& line& v1& v2& v3& v4& v1& v2& v3& v4 \\
    \hline \hline
    VGGNets& \textit{m}& 0.06& 0.05& 0.06& 0.09& 0.05& 0.06& 0.06& \textbf{0.12} \\
    & \textit{b}& 0.64& 0.67& 0.67& 0.64& 0.66& 0.66& \textbf{0.72}& 0.61 \\
    \hline
    ResNets& \textit{m}& 0.05& 0.08& 0.05& 0.08& \textbf{0.09}& 0.07& 0.07& 0.06 \\
    & \textit{b}& 0.39& 0.32& 0.44& \textbf{0.49}& 0.28& 0.38& 0.43& 0.54 \\
    \hline
    MobileNets& \textit{m}& 0.21& 0.21& 0.16& \textbf{0.22}& 0.19& 0.19& 0.18& 0.21 \\
    & \textit{b}& 0.23& 0.35& \textbf{0.44}& 0.39& 0.27& 0.38& 0.43& 0.41 \\
    \hline
    \end{tabular}
    \caption{\label{tab:linear}Linear regression analysis of \textit{Weight Skewness (WS)} and \textit{Prune Knee (PK)}. $m$ and $b$ are slope and bias of the linear equation $\textit{PK}=m\cdot \textit{WS}+b$, which can be solved by Equation (\ref{eq:m}) and (\ref{eq:b}). The data in bold highlights the highest value for each network architecture.}
    \end{table*}

Tab \ref{tab:pearson} shows the Pearson correlation coefficients between \textit{WS} and \textit{PK}, which is restricted to providing linear correlation and may not be applicable in situations where the correlation between variables is not strictly linear. However, the results presented in the table demonstrate a robust positive correlation that varies depending on the pruning methods, networks and hyper-parameters. MobileNets relatively have robuster and higher correlation values than VGGNets and ResNets. The linear relationship gets stronger with the increase of width for MobileNets while gets weaker with the increase of width for VGGNets. On the other hand, the pruning method proposed by Ye et al. \cite{ye2018rethinking} takes little advantages over that by Liu et al. \cite{liu2017learning} for VGGNets and MobileNets. However, the impact of pruning methods and hyper-parameters on this linear correlation for ResNets is unclear.

Tab \ref{tab:linear} shows the slope and bias of linear regression analysis between \textit{WS} and \textit{PK}. The pruning method of Ye et al. \cite{ye2018rethinking} gives similar results as that of Liu et al. \cite{liu2017learning}. In addition, MobileNet(width=1.00) gets the highest slope of the regression line while VGGNet(depth=16) has the largest bias.

\section{Discussion}
\label{sec:discussion}
Our results suggest that there is a significant positive relationship between \textit{WS} and \textit{PK}, which demonstrates that models with higher levels of channel sparsity are more likely to induce higher performance of pruning. This finding is consistent with the phenomenon observed in previous research which focus on developing various channel pruning methods.

The slope value obtained from the linear regression analysis (see Tab \ref{tab:linear}) indicates the pruning sensitivity on channel sparsity. MobileNets are relatively most sensitive to the change of channel sparsity. Together with Tab \ref{tab:pearson}, MobileNets also give the highest linear correlation coefficients. Therefore, it is safe to say that sparsifying the channels in MobileNets brings more benefits to pruning than VGGNets and ResNets.

The bias value represents pruning performance without channel sparsification. VGGNets have larger bias values than other two CNNs, which means that most channels in VGGNets can be safely pruned without introducing channel sparsification. One hypothesis for this is that VGGNets have too many parameters, most of which contributes little to the final classification.

Comparing the results between pruning methods, differences are negligible. It means pruning methods have little influence on the relationship between channel sparsity and pruning. Additionally, regular pattern is hardly found from the results between different hyper-parameters (width or depth) of a single network architecture, which indicates that the impact of hyper-parameters (width or depth) is still mysterious.

Overall, our findings provide support for the existence of a strong and positive linear relationship between channel sparsity and pruning performance. Further research could explore the underlying mechanisms that drive this relationship, and whether other interventions could also have effects on this relationship.

\section{Conclusion}
\label{sec:conclusion}
This paper proposes a novel metric to measure the channel sparsity in ASC systems, and provides a new way to examine the performance of neural network pruning. The experiment results demonstrate a strong and positive linear correlation between \textit{Weight Skewness (WS)} and \textit{Prune Knee (PK)}, which indicates that models with higher channel sparsity can bring about a larger compression rate without sacrificing too much accuracy. The choice of pruning method does little impact on this relationship. In comparison to VGGNets and ResNets, MobileNets experience greater advantages from channel sparsification when undergoing pruning.

\section{Acknowledgement}
The research project is supported partly by the National Natural Science Foundation of China (No: 62001038) and Gusu Innovation and Entrepreneurship Leading Talents Programme (No: ZXL2022472).


\bibliographystyle{IEEEbib}
\bibliography{reference}

\end{document}